\documentclass[10pt,conference]{IEEEtran}

\usepackage{cite}
\usepackage{todonotes}
\usepackage{amsmath,amssymb,amsfonts}
\usepackage{algorithmic}
\usepackage{graphicx}
\usepackage{textcomp}
\usepackage{booktabs} \usepackage{tabularx}
\usepackage{xcolor}
\usepackage{hyperref}
\usepackage{pgfplots}
\pgfplotsset{compat=1.18}
\usepgfplotslibrary{groupplots}
\usetikzlibrary{positioning,arrows.meta,calc}
\definecolor{opened}{HTML}{2A78D6}
\definecolor{merged}{HTML}{EB6834}
\definecolor{detfill}{HTML}{EEEDE8}
\definecolor{detline}{HTML}{8A897F}
\definecolor{llmfill}{HTML}{CDE2FB}
\definecolor{llmline}{HTML}{2A78D6}
\definecolor{textprimary}{HTML}{1A1A19}
\definecolor{textsecondary}{HTML}{5F5E58}
\definecolor{textmuted}{HTML}{8A897F}
\definecolor{gridline}{HTML}{E6E5DF}
\def\BibTeX{{\rm B\kern-.05em{\sc i\kern-.025em b}\kern-.08emT\kern-.1667em\lower.7ex\hbox{E}\kern-.125emX}}

\usepackage{array}

\makeatletter
\newcommand{\linebreakand}{%
  \end{@IEEEauthorhalign}
  \hfill\mbox{}\par
  \mbox{}\hfill\begin{@IEEEauthorhalign}
}
\makeatother
\begin{document}

\title{Hybrid LLM--Deterministic Pipelines for Enterprise-Scale Code Migration at Bloomberg}

\author{\IEEEauthorblockN{Mario Corchero}
\IEEEauthorblockA{\textit{Bloomberg}\\
London, United Kingdom \\
mcorcherojim@bloomberg.net}
\and
\IEEEauthorblockN{David Williams}
\IEEEauthorblockA{\textit{University College London}\\
London, United Kingdom \\
david.williams.22@ucl.ac.uk}
\and
\IEEEauthorblockN{L\'aszl\'o Kiss Koll\'ar}
\IEEEauthorblockA{\textit{Bloomberg}\\
London, United Kingdom \\
lkisskollar@bloomberg.net}
\and
\IEEEauthorblockN{Haowen Liu}
\IEEEauthorblockA{\textit{Bloomberg}\\
New York, United States \\
hliu746@bloomberg.net}
\linebreakand
\IEEEauthorblockN{Shivam Bairoliya}
\IEEEauthorblockA{\textit{Bloomberg}\\
London, United Kingdom \\
sbairoliya@bloomberg.net}
\and
\IEEEauthorblockN{Gala Malbasic}
\IEEEauthorblockA{\textit{Bloomberg}\\
London, United Kingdom \\
gmalbasic1@bloomberg.net}
\and
\IEEEauthorblockN{Serkan Kirbas}
\IEEEauthorblockA{\textit{Bloomberg}\\
London, United Kingdom \\
skirbas@bloomberg.net}
\and
\IEEEauthorblockN{Federica Sarro}
\IEEEauthorblockA{\textit{University College London}\\
London, United Kingdom \\
f.sarro@ucl.ac.uk}
}

\maketitle

\begin{abstract}
Large organizations carry a long tail of infrastructure migrations that are worth doing but rarely urgent enough to compete with feature work, because each one costs significant engineering time per repository.

% In this paper we report on ``Automated Uplift'', an effort at Bloomberg to find out whether hybrid LLM-deterministc pipelines can lower that cost enough to make such migrations viable. We ran two campaigns at opposite ends of the automation spectrum: a C++ build system migration from a proprietary internal tool to CMake, and a Python packaging migration to a dependency-isolated model. Both pipelines follow what we call a deterministic envelope: classification, context collection, and validation are deterministic, and the LLM performs a single, well-specified transformation. For the Python campaign, the migration context also includes runtime production data, such as interpreter startup logs, module usage traces, and the dependency versions actually deployed.

This paper reports on our industrial experience from Automated Uplift, a multi-month initiative applying LLM-assisted migration pipelines across thousands of Bloomberg repositories. 

We examine two distinct production campaigns: a heterogeneous Python packaging migration ($\approx$1,900 candidate repositories) and a uniform C++ build system migration ($\approx$6,700 candidate repositories
). Together, these campaigns generated 2,333 pull requests, resulting in 729 successful migrations. We detail our hybrid architectural solution which maximizes deterministic processing, confines the LLM to bounded file transformations within a pre-computed context, and incorporates runtime production telemetry (such as execution logs and usage traces). 
%Finally, we analyse dominant adoption bottlenecks, highlighting that verification and deployment effort by owning teams remain the primary constraints at fleet scale, and provide actionable lessons for engineering hybrid AI systems in enterprise environments.
%The two campaigns generated 2,333 pull requests. By the end of the observation period, 34.2\% of the CMake PRs and 10.5\% of the Python PRs had been merged, but in both campaigns most PRs (58\% and 68\%) had received no decision from the owning team. 

Our findings show that generating migrations turned out to be the easier part; the bottleneck was the cost, for owning teams, of reviewing, testing, and deploying them. These costs were lowered when changes could be verified directly in the pipeline before opening a PR (as with C++ compilation and test validation), and when the population was uniform enough for fixes to carry over from one repository to the next.

We conclude by distilling our experience and findings into actionable, transferable patterns for architecting hybrid LLM–deterministic pipelines and scaling code migrations across the enterprise.

%transferable lessons for engineering hybrid LLM--deterministic pipelines and conducting code migrations at enterprise scale.
\end{abstract}


\section{Introduction}
Software maintenance is a large and ongoing part of the cost of software. For long-lived systems, maintenance includes not only changes to application code but also keeping the surrounding infrastructure current. Build systems become obsolete, packaging formats evolve, frameworks are deprecated, and language runtimes advance, yet the production code persists, often for decades. Modernizing this substrate is unavoidable: unmigrated repositories accrue escalating maintenance costs, prevent performance improvements, and create operational fragility. 

The challenge is scale. A large enterprise may maintain thousands of repositories spanning multiple technology stacks, each with its own build configuration, dependency graph, and deployment topology. Migrating even one repository requires understanding its structure, runtime dependencies, production deployment context, and accumulated conventions. Multiplying these requirements across thousands of repositories far exceeds the manual effort that the benefit would justify.

Large Language Models (LLMs) have demonstrated impressive code transformation capabilities, and recent work has explored their application to migration tasks ranging from language translation~\cite{cai2025rustmap, liu2025llmigrate} to API migration~\cite{islam2026migratelib} to enterprise-scale change list generation~\cite{ziftci-2025-migrating}.

However, applying out-of-the-box LLMs directly to repository migration tasks breaks down at scale. Benchmark evaluations reveal sharp degradation with repository complexity: Li et al.~\cite{li2026repomodbench} report pass rates dropping from 91.3\% for repositories under 10,000 lines of code to 15.3\% for repositories exceeding 50,000 lines on RepoMod-Bench. Structurally, successful migration requires information absent from source code, such as which interpreter version runs in production, which dependency versions are deployed to which clusters, and which scripts are actually invoked at runtime versus dormant. 

Scale is not the only obstacle; priority is another. Like most large engineering organizations, Bloomberg has a record of executing company-wide migrations and keeping its codebase up to date: urgent fleet-wide patches, end-of-life runtimes, and deprecations with hard deadlines are funded, centrally coordinated, and tracked to completion. Below these high-profile cases sits a long tail of migrations that are worthwhile but never urgent enough to compete with feature work, such as moving off an earlier-generation build tool that still works or adopting a packaging model that reduces operational risk. Individually, each is low priority, but collectively they represent a significant opportunity to simplify and modernize the organization's infrastructure. The economics explain the backlog: when the migration cost per repository is high, only the most important ones clear the bar. This raises the question that motivates our work: \textit{if automation drastically lowers the cost of producing a migration, can an organization take on migrations that were previously not worth doing?}

Attempting to answer this question raises a more practical one: \textit{how does one build a reliable, cost-effective migration system that operates across thousands of diverse repositories, each with its own structural idiosyncrasies and production context?}

At Bloomberg, we call this kind of work \emph{uplift}: modernizing infrastructure at the repository level by moving repositories from earlier-generation tooling to current equivalents. Uplift does not change what the code does; it changes how it is built, packaged, deployed, and maintained, with the aim of lowering its future maintenance cost. Historically, a central team driving a low-priority uplift would announce it, publish instructions and an FAQ, and ask each owning team to perform the migration themselves; for example, the owners of the earlier-generation proprietary C++ build tool would ask every team to move to CMake. Each team then rediscovered the same edge cases independently. Our work improves that model: the central team encodes its instructions for a pipeline that generates the migration PRs, and owning teams review and deploy them.

This paper presents our industrial experience from Automated Uplift, a project at Bloomberg aimed at answering these questions. Over six months of campaigns, we developed and applied LLM-assisted migration pipelines at Bloomberg. We piloted across five technology domains, and in this paper we report on the two that grew into sustained campaigns. These campaigns differ on several key dimensions that matter most for automation: how uniform the source population is, how deep the validation feedback loop is (Section~\ref{sec:feedback-loops}), and therefore what role the LLM plays:

\begin{enumerate}
    \item \textbf{Python packaging migration} (``PyPackaging''): Migrating Python packages from a proprietary shared-environment packaging model to \emph{PyPackaging},\footnote{PyPackaging and CppBuild are pseudonyms for internal Bloomberg tools.} an internal dependency-isolated model in which each package declares and bundles its own dependencies. The source population is heterogeneous, and the pipeline's checks give only partial confidence in a change. This campaign targets $\approx$1,900 candidate repositories (with 294 PRs generated).
    \item \textbf{C++ build system migration} (``CppBuild-to-CMake''): Migrating C++ projects from \emph{CppBuild}, an internal proprietary build system, to CMake, including restructuring source trees, translating build configurations, and resolving dependency declarations. The source population is uniform, and the pipeline's checks give high confidence in a change. This campaign targets $\approx$6,700 candidate repositories (with 2,039 PRs generated).
\end{enumerate}

Together, the two campaigns target approximately 8,600 repositories and generated more than 2,300 pull requests. Both involve semantic adaptation beyond syntactic transformation: understanding what the code does in production, how it is deployed, and what it depends on at runtime. Both pipelines also share a common design principle that 
underlies our hybrid architecture: maximize deterministic processing and confine the LLM to file transformation within a fully characterized, pre-computed context. The overall pipeline is deterministic, and the LLM handles one bounded task within it. Moreover, we provide the LLM with ``migration context'' from runtime production data, such as execution logs, module usage traces, and cluster-specific dependency versions, cross-referencing AST-inferred imports against runtime-observed usage.

We structure our study around three research questions:

\begin{itemize}
    \item \textbf{RQ1: How should AI-assisted migration pipelines partition work between deterministic tooling and LLM-driven steps?} We look at how the work is split in our two pipelines and how that split changed as models improved during the study.
    \item \textbf{RQ2: What is the impact of incorporating runtime production data on migration accuracy?} We describe how production telemetry feeds the migration context and what changed once we added it, compared to relying on source code alone.
    \item \textbf{RQ3: What are the dominant bottlenecks in fleet-scale migration campaigns?}  We analyze the outcomes of more than 2,300 PRs across both campaigns to find where migrations stall once PRs are generated.
\end{itemize}

Over the study period, the two campaigns generated 2,333 PRs. Two months or more after being opened, 34.2\% of CppBuild-to-CMake PRs and 10.5\% of PyPackaging PRs had been merged, totaling 729 migrations. These rates can look low, but they should be read against the alternative: these are low-priority migrations that, without automation, would have stayed in each owning team's backlog, waiting to be prioritized. By the team's estimates, the merged changes alone represent well over a thousand person-days of manual migration work. Most unmerged PRs were not rejected but simply received no decision, and 34 of the 51 PyPackaging repositories where our PR was rejected later completed the migration themselves. Runtime production data proved essential for packaging migrations. Finally, the main constraint was not generating changes but the cost for owning teams to verify and deploy them, which was lowest when the pipeline could verify a change itself before opening a PR.

In summary, our contributions are:

\begin{enumerate}
    \item Exploring hybrid LLM-deterministic pipelines for enterprise-scale code migration, where deterministic phases handle classification, data collection, and validation, while LLMs perform creative transformation within a constrained scope (RQ1).
    \item Exploring the use of runtime-data-enriched migration context: augmenting LLM prompts with production telemetry, including interpreter startup logs, module usage traces, and cluster-specific dependency versions, reconciling statically inferred imports with runtime-observed module usage (RQ2).
    \item Exploring fleet-scale migration operations across two campaigns covering more than 2,300 pull requests, unveiling that developer engagement and PR adoption, not generation quality, is the primary operational constraint at scale for this kind of migration (RQ3).
    \item Sharing the lessons learned from our experience on engineering hybrid AI-assisted migration systems in enterprise environments.
\end{enumerate}

\section{Background}

This section provides context on the infrastructure our pipelines draw on and the concept of validation feedback loops that we use to compare the two campaigns.

\subsection{Organizational Context}
Bloomberg operates its own internal infrastructure for package management, build orchestration, deployment, and production telemetry. Packages are distributed through an internal binary package registry, each with its own versioning, dependency resolution, and deployment semantics. Production telemetry, including interpreter startup logs, module usage traces, and process-level resource consumption, is captured by a centralized observability platform. This provides a runtime view of how code actually executes in production. The infrastructure serves as a data source with no analog in open source or cloud native environments: the ability to observe, at fleet scale, which scripts run, which modules they import, and which dependency versions are loaded on which clusters. Section~\ref{sec:system-architecture} describes how the system incorporates this telemetry into the migration pipeline.

\subsection{Validation Feedback Loops} \label{sec:feedback-loops}
Every automated migration needs a mechanism to tell whether a generated change is correct. We call the checks that do this \emph{validation feedback loops}, and we describe their \emph{depth} as how far an error can travel between the step that introduces it and the first check that can catch it. In a \emph{shallow} pipeline, each step hands its output to a check that validates it before the next step runs, so the migration is verified as a series of short loops, all within the pipeline. CppBuild-to-CMake works this way: an extraction error is caught by schema validation, a template error by the build, and a behavioral change by running unit tests, before any PR is opened. In a \emph{deep} pipeline, intermediate checks cannot confirm that earlier steps were right, and the only complete check comes at the end. For PyPackaging, a package can build and pass smoke checks such as \texttt{--help}, yet a wrong dependency choice made at the start only surfaces when the owning team deploys and runs the application. When that happens, the migration must be reworked from the start. Section~\ref{sec:rq3-feedback-loop-depth} examines how this difference shaped the outcomes of the two campaigns.

\section{System Architecture} \label{sec:system-architecture}
This section describes the two pipelines. We start with the structure they share, then cover what is specific to each, and close with how the split between deterministic and LLM-driven work evolved during the project. Table~\ref{tab:pipeline_comparison} summarizes the differences.

Progress on both campaigns is recorded in a centralized initiative tracker. Each target repository appears as a task assigned to its owning team, and the task's status updates automatically as the migration lands in the repository, so both the central team and owning teams can see where each migration stands. 

\subsection{Hybrid LLM-Deterministic Pipelines for Code Migration} \label{sec:deterministic-envelope}
\begin{figure}[!t]
\centering
\begin{tikzpicture}[
  font=\scriptsize, text=textprimary,
  stage/.style={draw, rounded corners=2pt, line width=0.6pt, minimum width=2.3cm, minimum height=0.62cm, align=center, font=\scriptsize\bfseries},
  det/.style={stage, fill=detfill, draw=detline},
  llm/.style={stage, fill=llmfill, draw=llmline},
  note/.style={anchor=west, align=flush left, text width=4.45cm, font={\scriptsize\hyphenpenalty=10000 \exhyphenpenalty=10000 }, text=textsecondary},
  flow/.style={-{Stealth[length=4pt]}, line width=0.6pt, draw=detline},
  node distance=0.32cm,
]
\node[det] (s1) {1. Classify \& collect};
\node[det, below=of s1] (s2) {2. Render prompt};
\node[llm, below=of s2] (s3) {3. Execute};
\node[det, below=of s3] (s4) {4. Validate};
\node[llm, below=of s4] (s5) {5. Describe};
\node[stage, draw=detline, dashed, fill=white, font=\scriptsize, below=of s5] (out) {Open PR};

\foreach \a/\b in {s1/s2, s2/s3, s3/s4, s4/s5, s5/out} \draw[flow] (\a) -- (\b);
\draw[flow, draw=llmline] (s4.west) -- ++(-0.35,0) |- node[pos=0.25, left, font=\scriptsize, text=textsecondary, align=right] {retry\\up to 3} (s3.west);

\node[note, right=0.3cm of s1] {Archetype from a static registry; package metadata, AST imports, build config; runtime telemetry (PyPackaging only)};
\node[note, right=0.3cm of s2] {Archetype-specific template; same context, same prompt};
\node[note, right=0.3cm of s3] {PyPackaging: LLM edits the repository\\CMake: LLM extracts JSON; templates write \texttt{CMakeLists.txt}};
\node[note, right=0.3cm of s4] {Rules from past failures; CMake: schema, build, unit tests; PyPackaging: packaging, \texttt{--help} diff};
\node[note, right=0.3cm of s5] {Read-only; summarizes evidence and remaining warnings};
\node[note, right=0.3cm of out] {Reviewed in batches by the central team before the owning team is asked to act};

\node[det, minimum width=0.35cm, minimum height=0.25cm, anchor=west] (k1) at ($(out.south west)+(0,-0.55)$) {};
\node[anchor=west, font=\scriptsize, text=textsecondary] at (k1.east) {Deterministic};
\node[llm, minimum width=0.35cm, minimum height=0.25cm, anchor=west] (k2) at ($(k1.east)+(1.9,0)$) {};
\node[anchor=west, font=\scriptsize, text=textsecondary] at (k2.east) {LLM};
\end{tikzpicture}
\caption{The pipeline shared by both campaigns. Grey stages are deterministic; blue stages call the LLM.}
\label{fig:pipeline}
\end{figure}
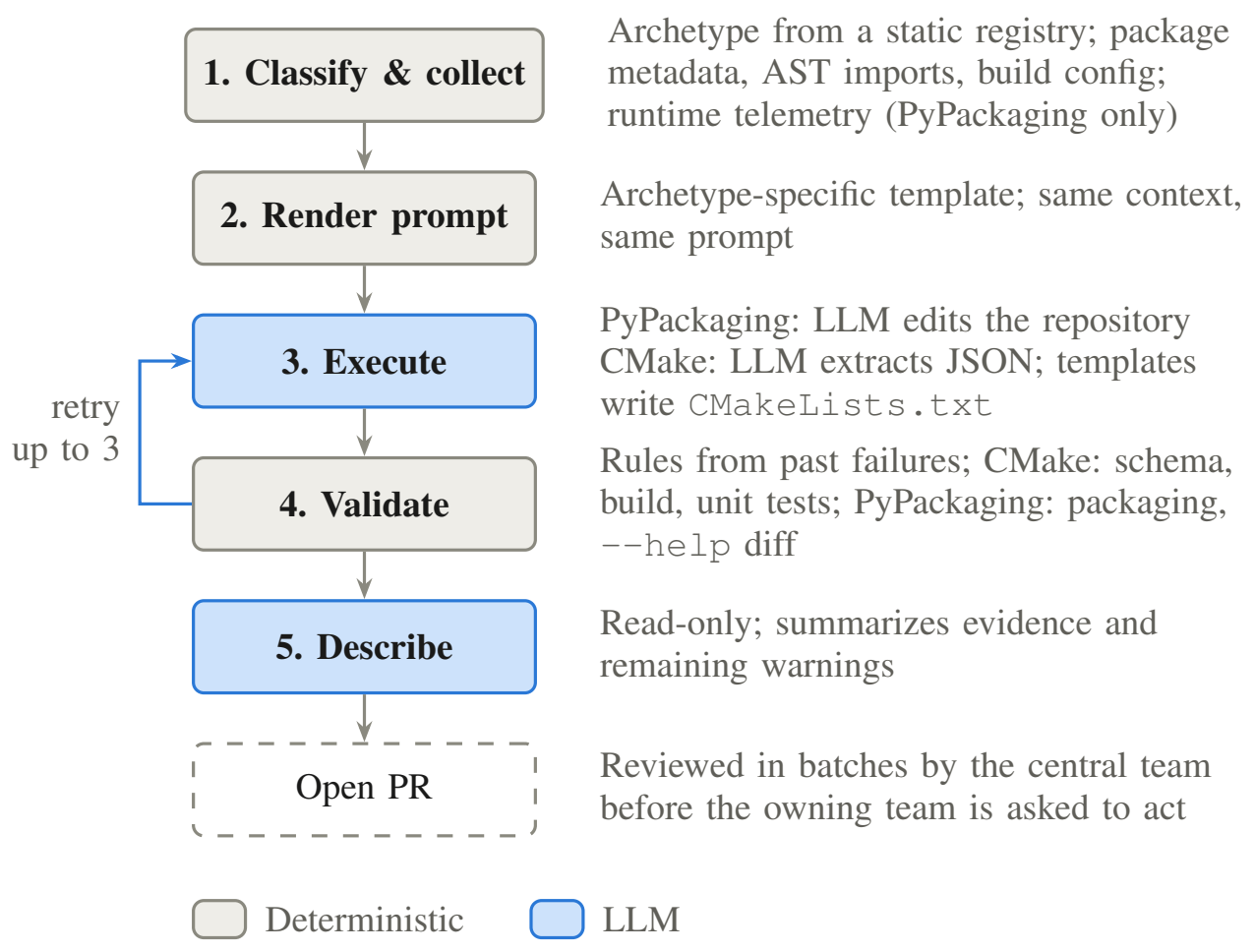

Both pipelines run the same five stages for each repository (as depicted in Figure~\ref{fig:pipeline}):
\begin{enumerate}
    \item \textbf{Classify and collect} (deterministic). Assign the repository to an \emph{archetype}, a class of repositories that share a structure and therefore the same migration approach (e.g., Python scripts with no packaging metadata), and assemble the migration context: metadata, build configuration, source analysis and, for PyPackaging, runtime telemetry.
    \item \textbf{Render prompt} (deterministic). Turn the context into a prompt with an archetype-specific template. The same context always produces the same prompt.
    \item \textbf{Execute} (LLM). Perform the migration, either by editing files (PyPackaging) or by extracting structured data (CMake).
    \item \textbf{Validate} (deterministic). Check the result against rules derived from past failures. If a check fails, re-invoke the LLM with the errors provided as additional context, up to three times.
    \item \textbf{Describe} (LLM). A read-only invocation writes the PR description from the migration evidence, including any warnings left after validation.
\end{enumerate}

The principle behind this structure is intuitive: anything that must be reproducible or auditable across the fleet is deterministic, and the LLM only handles the part that varies from one repository to the next. Validation is never delegated to the model. Each migration runs in its own container, and the central team reviewed generated PRs in batches before handing them to owning teams. This step was meant to capture clear mistakes before even asking owning teams to review.

\begin{table}[!t]
\renewcommand{\arraystretch}{1.15}
\caption{Core Differences Across the Two Campaigns}
\label{tab:pipeline_comparison}
\centering
\footnotesize
\begin{tabularx}{\columnwidth}{@{}>{\raggedright\arraybackslash}p{0.24\columnwidth} >{\raggedright\arraybackslash}X >{\raggedright\arraybackslash}X@{}}
\toprule
 & \textbf{CppBuild-to-CMake} & \textbf{PyPackaging} \\
\midrule
Source population & Uniform (CppBuild makefiles) & Heterogeneous (18 categories) \\
LLM role & Extracts structured data; templates write the code & Writes the migration directly \\
Runtime telemetry & Not needed & Central input \\
Pre-PR validation & Schema, build, unit tests & Configuration rules, packaging, \texttt{--help} diff \\
Feedback loop & Shallow & Deep \\
Primary model & Claude Sonnet 4.5 & Claude Opus 4.6 \\
Mean agent run & $\sim$13 turns, 48\,s & $\sim$63 turns, 7.2\,min \\
\bottomrule
\end{tabularx}
\end{table}

\subsection{PyPackaging: The LLM Writes the Migration} \label{sec:architecture_pypackaging}
PyPackaging input candidates vary widely, from bare scripts with no build system to mixed C++/Python repositories, so the LLM has to write the migration itself: the packaging configuration, source layout changes, and build updates.

\textbf{Archetypes.} Each package is classified by looking it up in a precomputed registry built from the production asset registry, the package registry and the initiative tracker. We had tried LLM-based classification in an earlier campaign and dropped it: the same package could land in different archetypes on different runs, failures were harder to debug, and calling a model for every candidate in the fleet was wasteful. The 1,880 PyPackaging tasks fall into 18 categories. Three categories (covering 337 tasks) have dedicated prompt templates (500--600 lines each), data collectors, and validation rules. Two cover single-script packages: \emph{current build, no metadata} (already on the current build system, missing only packaging metadata) and \emph{proprietary packaging only} (shipped through the proprietary shared-environment model and packaged from scratch). The third, \emph{broken existing packaging}, covers repositories with a partial or misconfigured setup. Everything else gets a generic prompt that produces a starting-point PR, which is labeled as such for the owning team. Archetypes with $\approx$20 or fewer repositories did not justify developing a custom template.

\textbf{Runtime data.} \label{sec:PyPackaging-runtime-data} The migration context draws on 12 data sources, which fall into four groups: runtime telemetry from the observability platform (interpreter startup logs and module usage traces), package metadata, AST-based source analysis and build configuration. The key use is dependency resolution. Imports found by AST scanning are cross-referenced with modules observed at runtime. Dependencies seen in both are \emph{validated} and given the version available for the application in the deployed environment. Dependencies seen only in the source may be conditional, test-only, or dead, so they are passed to the LLM without a pin. Modules seen only at runtime are usually transitive and are left out.

\textbf{Import-to-package resolution.} Module usage traces and source analysis both tell us which modules an application imports, but a migration has to declare the canonical package names that provide them, and in Python the two names often differ: \texttt{PIL} comes from \texttt{pillow}, \texttt{yaml} from \texttt{PyYAML}, and a namespace such as \texttt{google} is split across many packages. A wrong choice still builds and passes our checks, and only fails once the application runs (Section~\ref{sec:feedback-loops}). Rather than resolving names separately for each repository, we built one fleet-wide mapping from the modules observed in production to the packages that provide them, using public import-to-package metadata~\cite{condaforge-cfgraph} for open source packages and the internal package index for internal ones.

\textbf{Execution and validation.} The agent runs with a fixed set of file and shell tools, no network access, and is locked to the repository directory. After it finishes, 7 rules check for oversights we had seen repeatedly in prior experiments, such as legacy configuration left in place or files copied instead of linked. The pipeline also performs a simple import validation for CLI scripts by invoking them with \texttt{--help} before and after the migration, and compares the packaged layout with the current deployment. These checks catch packaging mistakes but not runtime failures, which is what makes this feedback loop deep (Section~\ref{sec:feedback-loops}).

\subsection{CppBuild-to-CMake: LLM Extracts, Templates Write}
CppBuild makefiles follow a well-defined format, so the LLM never writes build files. It reads the makefile and extracts sources, libraries, dependencies, flags and custom rules into a JSON object validated against a schema. Templates then generate the \texttt{CMakeLists.txt}. A wrong build file is therefore either an extraction error, fixed by tightening the schema or prompt, or a template bug, fixed once for every repository. Validation checks the schema, builds the project, and runs its unit tests. No runtime telemetry is needed, since the build configuration does not depend on runtime behavior.

The CMake backlog splits into three archetypes (shown in Table~\ref{tab:archetype_families}). One of them, company-defined build configurations, follows a fully deterministic path: a dedicated generator converts them with no LLM involvement at all.

\begin{table}[!t]
\renewcommand{\arraystretch}{1.15}
\caption{CppBuild-to-CMake PR Generation by Archetype}
\label{tab:archetype_families}
\centering
\footnotesize
\setlength{\tabcolsep}{3.5pt}
\begin{tabularx}{\columnwidth}{@{}>{\raggedright\arraybackslash}X r r r@{}}
\toprule
\textbf{Archetype} & \textbf{Tasks} & \textbf{PRs} & \textbf{Rate} \\
\midrule
Company-Defined Build Configuration & 1,877 & N/A$^{\mathrm{a}}$ & N/A \\
Team-Defined Build Configuration & 2,889 & 1,283 & 44.4\% \\
Code Generation & 1,355 & 854 & 63.0\% \\
\bottomrule
\end{tabularx}
\par\smallskip\noindent\parbox{\columnwidth}{$^{\mathrm{a}}$A deterministic uplift tool is available due to the high level of convergence resulting from the strict standardization applied on those projects prior to this uplift effort. LLM-driven uplifts were therefore not the major avenue of uplift for those projects.}
\end{table}

\subsection{How the Boundary Moved} \label{sec:expanding-trust-boundary} \label{sec:trust-boundary-implications}
Over the project, we updated the underlying model from Claude Sonnet 4.5 to Opus 4.6, and the split between deterministic and LLM-driven work moved with it, in both directions. As models improved, we relaxed some deterministic pre-processing, such as AST-based dependency handling that compensated for earlier models' unreliable import analysis, and replaced concise instructions with richer reference documents. The limiting factor became the precision of our documentation, rather than the models' capabilities. In the other direction, we moved archetype classification for PyPackaging tasks from the LLM in our early experiments to a registry for the reasons above (Section~\ref{sec:architecture_pypackaging}), and kept all validation deterministic throughout, with new validations only ever added. Our working rule became: expand the LLM's role in generating changes, but not in verifying them.

\section{Evaluation} \label{sec:evaluation}
The project started in October 2025. After three months of experimentation, both campaigns ran from January 2026. We report on all campaign PRs created up to 30 April 2026 and measure their outcomes as of 30 June 2026, when every PR was at least two months old. We exclude test PRs we created on personal and sandbox repositories, and PRs the bot closed itself, mostly to regenerate them. Each remaining PR is labeled \emph{merged}, \emph{rejected} (closed unmerged by a developer), or \emph{no decision} (no developer decision by 30 June: still open, or closed by automation such as stale-PR bots).

\subsection{RQ1: Partitioning Work Between Deterministic and LLM Steps}
Classification, context collection, prompt rendering, and validation checks are deterministic. The LLM is called at three points: to execute the migration, to remediate after a failed check, and to write the PR description. Around these calls sit the registry, 12 data collectors, 7 validation rules, the renderer, the orchestrator, PR lifecycle management, and a dashboard. In code, the PyPackaging migration tool has about 4,300 lines of deterministic Python against 230 lines of LLM interface, plus about 3,600 lines of prompt templates and reference documentation. We give these counts as a description of where the design decisions sit, not as a measure of effort.

Table~\ref{tab:pr-generation-results} shows how often the pipeline produced a PR for the three PyPackaging archetypes with dedicated templates. The simplest archetype, \emph{current build, no metadata}, reached 80.2\%. The other two, \emph{proprietary packaging only} and \emph{broken existing packaging}, reached about 55--57\%. Most failures were deliberate: the pipeline declines to open a PR when data collection shows something it cannot migrate safely, such as unresolvable dependencies or several packages in one repository.

\begin{table}[!t]
\renewcommand{\arraystretch}{1.15}
\caption{PyPackaging PR Generation by Archetype}
\label{tab:pr-generation-results}
\centering
\footnotesize
\setlength{\tabcolsep}{3.5pt}
\begin{tabularx}{\columnwidth}{@{}>{\raggedright\arraybackslash}X r r r@{}}
\toprule
\textbf{Archetype} & \textbf{Tasks} & \textbf{PRs} & \textbf{Rate} \\
\midrule
Current Build, No Metadata & 96 & 77 & 80.2\% \\
Proprietary Packaging Only & 186 & 105 & 56.5\% \\
Broken Existing Packaging & 55 & 30 & 54.5\% \\
\bottomrule
\end{tabularx}
\par\smallskip\noindent\parbox{\columnwidth}{The first two rows are single-script packages: one already uses the current build system and needs only packaging metadata; the other is shipped through the proprietary shared-environment model and is packaged from scratch. The third covers repositories with a partial or misconfigured packaging setup. Tasks are matched to PRs by tags in the initiative tracker.}
\end{table}

The boundary shifts described in Section~\ref{sec:expanding-trust-boundary} show up in agent run times. PyPackaging PRs generated with Sonnet 4.5 averaged 4.1 minutes; later PRs with Opus 4.6 handling more complex archetypes averaged 7.7 minutes. The archetype mix also changed during this period, so we cannot clearly separate the effect of the model from that of the more challenging tasks.

The partition that worked best across our experiments keeps classification, context assembly, and validation deterministic and tasks the LLM with a single, well-specified transformation. Where the source format is regular enough, as for one CMake archetype, the LLM can be removed altogether.

\subsection{RQ2: Runtime Production Data}
As we did not run an explicit ablation study comparing pipelines with and without runtime data, we provide qualitative evidence, informed by the team's experience in the pilot studies leading up to the two migration campaigns. Before runtime data was integrated, the PyPackaging pipeline relied on AST scanning alone. In adding it, we observed four key changes. First, conditional and test-only imports stopped being treated as hard requirements. Second, dynamically loaded modules, for instance through plugins, became visible, and the pipeline flagged repositories that needed more care. Third, startup logs showed which interpreter version actually ran in production and which scripts were actually executed, which provides insight into the target version and entry points to preserve. Finally, cluster-specific versions removed a class of failures in which a declared version range was valid on one cluster but broke on another, since production clusters run different dependency versions. None of this information can be retrieved from source code alone. Thus, we consider runtime data essential for packaging migrations, but further work is still needed to quantify its effect.

\subsection{RQ3: Bottlenecks in Fleet-Scale Campaigns}
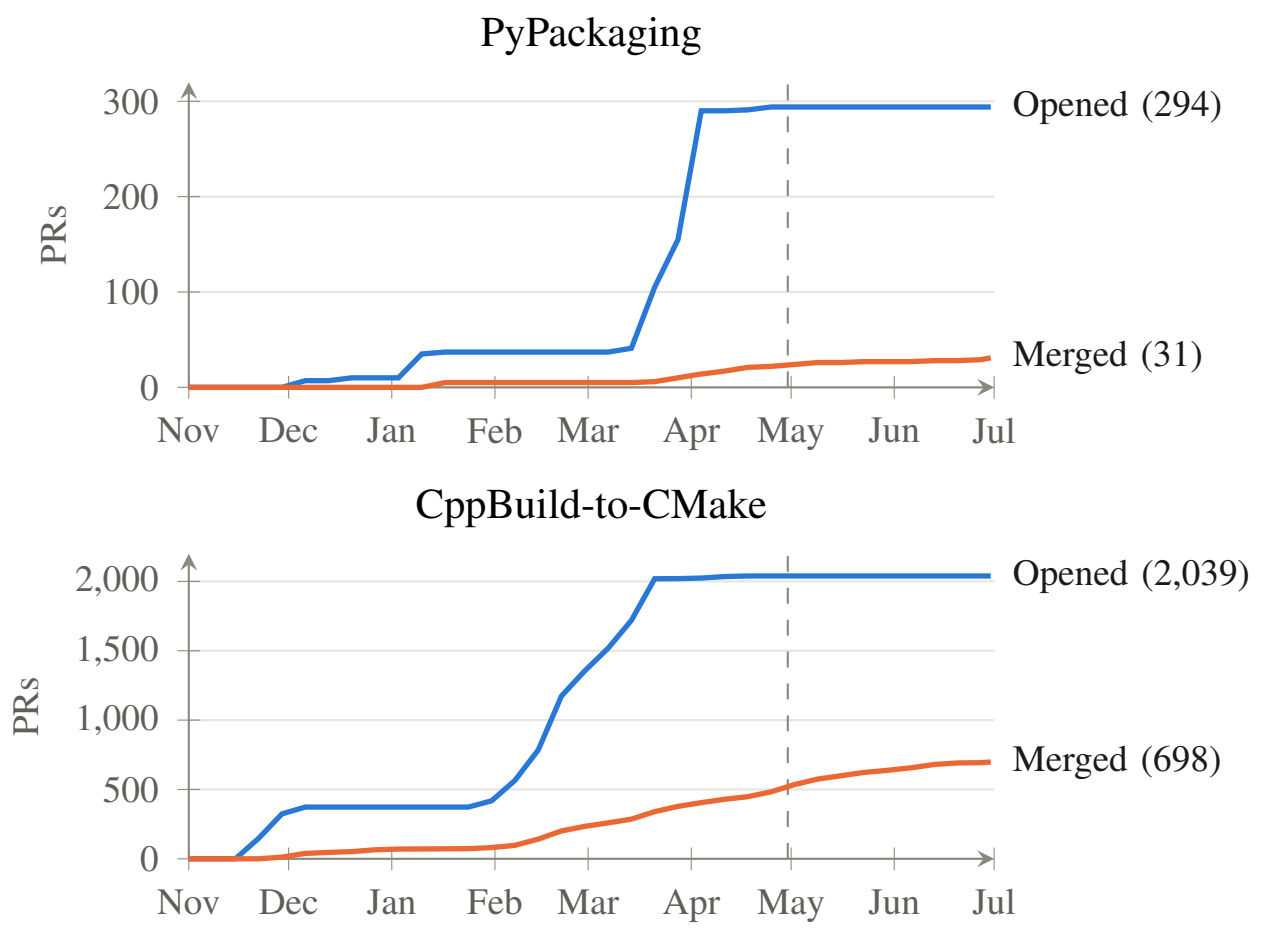
\begin{figure}[!t]
\centering
\begin{tikzpicture}
\begin{groupplot}[
  group style={group size=1 by 2, vertical sep=1.1cm},
  width=0.78\columnwidth, height=3.6cm,
  scale only axis=false,
  xmin=0, xmax=242, ymin=0,
  xtick={0,30,61,92,120,151,181,212,242},
  xticklabels={Nov,Dec,Jan,Feb,Mar,Apr,May,Jun,Jul},
  tick label style={font=\scriptsize, text=textsecondary},
  title style={font=\footnotesize, yshift=-1ex},
  ylabel={PRs}, ylabel style={font=\scriptsize, text=textsecondary},
  axis lines=left, axis line style={textmuted},
  tick style={textmuted},
  ymajorgrids, grid style={gridline},
  clip=false,
  scaled y ticks=false,
  yticklabel style={/pgf/number format/.cd, 1000 sep={,}, assume math mode=true},
]
\nextgroupplot[title={PyPackaging}, ymax=320, ytick distance=100]
\draw[textmuted, dashed, thin] (axis cs:180,0) -- (axis cs:180,320);
\addplot[opened, line width=1pt] coordinates {(0,0) (7,0) (14,0) (21,0) (28,0) (35,7) (42,7) (49,10) (56,10) (63,10) (70,35) (77,37) (84,37) (91,37) (98,37) (105,37) (112,37) (119,37) (126,37) (133,41) (140,105) (147,155) (154,290) (161,290) (168,291) (175,294) (182,294) (189,294) (196,294) (203,294) (210,294) (217,294) (224,294) (231,294) (238,294) (241,294)};
\addplot[merged, line width=1pt] coordinates {(0,0) (7,0) (14,0) (21,0) (28,0) (35,0) (42,0) (49,0) (56,0) (63,0) (70,0) (77,5) (84,5) (91,5) (98,5) (105,5) (112,5) (119,5) (126,5) (133,5) (140,6) (147,10) (154,14) (161,17) (168,21) (175,22) (182,24) (189,26) (196,26) (203,27) (210,27) (217,27) (224,28) (231,28) (238,29) (241,31)};
\node[anchor=west, font=\scriptsize, text=textprimary] at (axis cs:242,294) {Opened (294)};
\node[anchor=west, font=\scriptsize, text=textprimary] at (axis cs:242,31) {Merged (31)};
\nextgroupplot[title={CppBuild-to-CMake}, ymax=2200, ytick distance=500]
\draw[textmuted, dashed, thin] (axis cs:180,0) -- (axis cs:180,2200);
\addplot[opened, line width=1pt] coordinates {(0,0) (7,0) (14,0) (21,148) (28,324) (35,373) (42,373) (49,373) (56,373) (63,373) (70,373) (77,373) (84,373) (91,418) (98,565) (105,784) (112,1174) (119,1355) (126,1518) (133,1719) (140,2019) (147,2020) (154,2024) (161,2035) (168,2039) (175,2039) (182,2039) (189,2039) (196,2039) (203,2039) (210,2039) (217,2039) (224,2039) (231,2039) (238,2039) (241,2039)};
\addplot[merged, line width=1pt] coordinates {(0,0) (7,0) (14,0) (21,0) (28,12) (35,39) (42,46) (49,52) (56,65) (63,70) (70,71) (77,72) (84,73) (91,81) (98,97) (105,142) (112,201) (119,234) (126,259) (133,286) (140,340) (147,378) (154,405) (161,429) (168,448) (175,483) (182,535) (189,575) (196,599) (203,623) (210,638) (217,656) (224,680) (231,691) (238,693) (241,698)};
\node[anchor=west, font=\scriptsize, text=textprimary] at (axis cs:242,2039) {Opened (2,039)};
\node[anchor=west, font=\scriptsize, text=textprimary] at (axis cs:242,698) {Merged (698)};
\end{groupplot}
\end{tikzpicture}
\caption{Cumulative PRs opened and merged per campaign (November 2025 to end of June 2026). Only PRs opened up to 30 April (dashed line) are included; merges are counted up to 30 June.}
\label{fig:merge_curves}
\end{figure}

Table~\ref{tab:campaign_metrics} shows the outcomes for all PRs generated across both campaigns. The most common outcome in both campaigns was no developer decision at all: 67.7\% of PyPackaging PRs and 57.7\% of CMake PRs were neither merged nor rejected by the end of the observation period. This result demonstrates that generating PRs was not the main constraint, but rather convincing owning teams to review, test, and deploy them was. Figure~\ref{fig:merge_curves} shows the same point over time. PRs were generated in a few short batches, while merges arrived at a steady pace that did not follow them: in CppBuild-to-CMake, merges were still rising at a constant rate two months after the last PR was opened.

\begin{table}[!t]
\renewcommand{\arraystretch}{1.15}
\caption{Campaign Outcomes (PRs created by 30 April 2026, outcomes on 30 June 2026)}
\label{tab:campaign_metrics}
\centering
\footnotesize
\begin{tabularx}{\columnwidth}{@{}>{\raggedright\arraybackslash}X r r@{}}
\toprule
 & \textbf{PyPackaging} & \textbf{CppBuild-to-CMake} \\
\midrule
Total PRs & 294 & 2,039 \\
Merged & 31 (10.5\%) & 698 (34.2\%) \\
Rejected & 64 (21.8\%) & 165 (8.1\%) \\
No decision & 199 (67.7\%) & 1,176 (57.7\%) \\
\bottomrule
\end{tabularx}
\end{table}

\subsubsection{Feedback Loop Depth and Population} \label{sec:rq3-feedback-loop-depth}
The two campaigns also diverge. CMake PRs were merged at more than three times the PyPackaging rate, and rejected at less than half the rate. We attribute much of this to the depth of the feedback loop (Section~\ref{sec:feedback-loops}). A CMake PR arrives having been compiled and tested, so a reviewer is mostly checking a build file that is already known to work. A PyPackaging PR has only passed packaging and smoke checks; whether it works is only known after the owner deploys it, so the owner carries the risk and is more likely to reject it or wait. The populations also differ. CMake projects fall into a few similar archetypes, so fixes made for one repository apply to many. In contrast, PyPackaging candidates are spread over 18 archetypes with many one-off cases. Other factors point the same way: the CMake campaign started earlier, build-system changes may be seen as lower risk, and a part-time engineer was designated to support reviews. With only two campaigns, we cannot investigate the effects of these individual factors in depth.

\subsubsection{Impact Beyond Merges} \label{sec:rq3-impact}
Merge rate undercounts impact. In 34 of the 51 PyPackaging repositories where our PR was rejected, the owning team went on to land the migration themselves, and 14 of their PRs reused most of our changes. Developers told us that a common reason for closing and re-creating a PR was that they could not push to the bot's branch. One wrote: ``Yes, the PR was very useful. My main feedback was that one of the changes broke some unit tests. It would have been nice if the bot was able to identify any run/fix any tests before posting the PR. I tried to have [an AI coding tool] post the changes to the PR, but it got confused and ended up creating a new PR, so I just merged that one.''
This feedback points to two improvements we plan to incorporate in future versions of the PyPackaging campaign: letting owning teams push changes directly to the generated PR, so that they can refine it instead of starting over, and running the project's own tests before a PR is opened.

\subsubsection{Effort}
LLM inference costs were negligible compared with the engineering effort of building and operating the pipelines. Domain experts estimate a manual migration at about 2 person-days per CppBuild-to-CMake task and 1--2 person-days per PyPackaging task. These estimates come from the team that built the system and should be read with that in mind.

\section{Discussion}
\subsection{Lessons Learned}
Six months of operation yielded lessons that we believe transfer to other organizations undertaking AI-assisted code migration at scale.

\textbf{Lesson 1: PR generation is no longer the bottleneck; receiving and validating the changes is.} The system produced migration PRs much faster than owning teams could absorb them. By the end of the observation period, 10.5\% of PyPackaging PRs and 34.2\% of CppBuild-to-CMake PRs had been merged, while 67.7\% and 57.7\% had received no developer decision at all. Further improvements to generation would not have changed these numbers much. What limited them was the cost to the owning team of reviewing, testing, and deploying each change, and how much they trusted that it would work. We therefore expect the next significant gains to come from reducing that cost: running more of the validation inside the pipeline, giving reviewers stronger evidence that a change is correct, and making it easy for teams to iterate on a generated PR rather than starting again.

The high rate of PRs with no decision is also partly a consequence of the low priority of these campaigns. Neither carries a deadline or a security driver, and the receiving team bears the review effort and deployment risk while the benefit is diffuse and long-term. Lowering the cost of \emph{producing} a migration does not lower the cost of \emph{absorbing} it, and for non-urgent work that absorption cost is what developers weigh. A low direct merge rate therefore reflects how teams prioritize unsolicited, non-urgent changes, not only the quality of the generated change itself.

\textbf{Lesson 2: Archetype decomposition was among the most impactful architectural decisions.} Classifying migration targets into archetypes enabled three capabilities that a one-size-fits-all approach would struggle with. First, it allowed per-archetype optimization, with each archetype receiving a dedicated prompt template, data collection strategy, and validation rule set. Second, it made success rates interpretable: the 80\% rate for the \emph{current build, no metadata} archetype describes a well-characterized group of repositories rather than an average across heterogeneous targets. Third, it supported prioritization, as knowing which archetypes were more automatable and which were more complex or blocked let the team sequence work by expected return. The pattern applied to both campaigns.

\textbf{Lesson 3: Source code analysis alone is insufficient for many packaging migrations.} We did not expect how much runtime data would matter. The accuracy improvement from incorporating production telemetry, including startup logs, module usage traces, and cluster-specific dependency versions, was large enough that we consider runtime data a critical input for packaging migrations, although we have not yet confirmed this with a controlled ablation (Section~\ref{sec:threats}). Source code analysis cannot determine which imports are exercised in production, which dependency versions are deployed to each cluster, or which scripts are actually invoked. The three-way dependency validation scheme (Section~\ref{sec:PyPackaging-runtime-data}) provides a principled mechanism, but the broader point generalizes: any migration system that operates on source code without runtime context works with an incomplete picture. This extends the insights from Li et al.~\cite{li2026environment} regarding ``Environment-in-the-Loop'' migration from the academic setting to an industrial-scale system.

\textbf{Lesson 4: Models improve faster than pipeline engineering.} Over six months, the underlying LLM progressed from Sonnet 4.5 to Opus 4.6, with substantial capability improvements. Deterministic workarounds that required weeks of engineering (e.g., the AST-based dependency pre-processing that compensated for earlier models’ unreliable import analysis) became less critical as newer models handled these cases directly. In hindsight, this creates a design tension. Hardcoding too much deterministic pre-processing creates technical debt when a newer model handles the task natively. Delegating too aggressively to the LLM creates fragility when the model’s behavior changes between versions. Our heuristic: expand the LLM’s role in content generation but not in verification (Section~\ref{sec:trust-boundary-implications}). 

Two practical consequences followed. First, we started revisiting deterministic pre-processing with each model upgrade: when a new model handled a case natively, we kept the deterministic step only if it also made the pipeline more reproducible or easier to debug. Second, the work moved from code to documentation: with more capable models, the limiting factor became how precise our instructions and reference documents were, not what the model could do. Verification stayed out of this trade-off; we never moved a validation rule into the LLM, because an error that slips through at fleet scale could harm hundreds of repositories.

\textbf{Lesson 5: Feedback loop depth determines the automation ceiling.} The CMake and PyPackaging campaigns revealed this structural factor clearly. CppBuild-to-CMake is a format translation where compilation success is a strong proxy for correctness: a shallow feedback loop that enables high automation confidence. PyPackaging transforms heterogeneous source formats that require runtime verification through the full deployment lifecycle, creating a deep feedback loop that caps automation confidence. Teams planning AI-assisted migration campaigns should assess feedback loop depth \textit{before} investing in pipeline engineering.

\textbf{Lesson 6: Impact exceeds merge rate.} A follow-up on the 51 repositories where our PyPackaging PRs were rejected found that 34 of the owners later handled the migration themselves, in 14 cases reusing most of our changes. Developers who closed and re-created PRs told us the generated PRs were useful as starting points. This has design implications: PR descriptions, commit messages, and inline comments should be optimized for readability and adaptability by a developer who intends to re-implement the changes. Systems that measure success solely by merge rate undercount their actual impact.

\textbf{Lesson 7: The long tail resists full automation.} Archetypes affecting $\approx$20 or fewer repositories do not justify dedicated automation. The engineering cost of handling each additional edge case exceeds the manual effort saved. For these cases, a starting-point PR that developers refine is the right answer. Recognizing this threshold early prevents over-investment in diminishing-return automation and redirects effort toward the adoption and landing challenges identified in Lesson 1.

\subsection{Threats to Validity} \label{sec:threats}
\textbf{Internal validity.} The most significant threat is the absence of controlled experiments. Pipeline code, prompt templates, validation rules, and model versions all changed during the evaluation period, often concurrently, making it difficult to attribute observed improvements to any single factor. Success rate improvements between January and April 2026 may reflect model capability gains, pipeline improvements, or selection bias (easier archetypes attempted first, potentially inflating early success rates such as the 80\% \emph{current build, no metadata} figure in Table~\ref{tab:pr-generation-results}). All qualitative judgments, including archetype definitions, cost estimates, and the interpretation of runtime data’s importance, come from the same team that built the system. Independent evaluation would mitigate this experimenter bias.

The evaluation lacks a controlled experiment comparing migration outcomes with and without runtime data enrichment. The claim that runtime data is critical (Lesson 3) rests on qualitative engineering judgment and the architectural argument in Section~\ref{sec:PyPackaging-runtime-data}, not on a quantitative ablation study. An experimental design comparing AST-only versus AST+runtime-log pipelines on the same task population would provide stronger evidence, but its cost, and the fact that the campaigns targeted live products owned by real teams, made us discard this option.

Finally, the model evolution during the study period makes our findings time-bound. The specific success rates and failure modes reported here reflect a particular configuration of model capabilities (Sonnet 4.5 and Opus 4.6 as of April 2026). The most fragile findings are the specific quantitative results: the 80\% \emph{current build, no metadata} generation rate and the split between deterministic and LLM-driven components will likely shift as models improve. The general principles of using hybrid pipelines, runtime data enrichment, archetype decomposition, and bounded autonomy are intended to be more durable, but their specific parameterization will need ongoing calibration.

\textbf{External validity.} The findings come from a single organization with proprietary packaging and build systems, organization-specific deployment patterns, and a particular culture regarding code ownership and PR review. While the scale is substantial ($\approx$8,600 candidate repositories, 2,333 PRs), all observations occur within a single organizational context. Organizations with monorepo architectures will not face the PR landing problem in the same form, as Google's reported experience shows~\cite{nikolov-2025-migrating, ziftci-2025-migrating}, where a single change can atomically update all affected code.
The general architectural patterns (hybrid pipelines, runtime data enrichment, archetype classification) are more likely to generalize than the specific implementations.

\textbf{Construct validity.} We report multiple metrics (generation success rate, merge rate, and decision outcomes) but none cleanly separates generation quality from developer engagement. A PR may go unmerged because the receiving team lacks bandwidth or disagrees with the migration priority, not because of a technical deficiency. The repositories that closed our PR and later landed the migration themselves (Section~\ref{sec:rq3-impact}) illustrate the ambiguity: technically rejected but functionally successful. We did not conduct a formal user study of developer experience. This pattern merits broader qualitative investigation through interviews or surveys. Understanding why developers chose to re-implement rather than merge, and what changes to PR format or migration approach would increase direct merge rates, requires structured user research beyond the ad hoc feedback we collected. The manual migration cost estimates (2 person-days per CppBuild-to-CMake task, 1--2 per PyPackaging task) come from expert judgment rather than timed measurements. A rigorous cost comparison would require timing a sample of manual migrations and computing fully-loaded cost equivalents.

\textbf{Conclusion validity.} The paper’s authors designed, built, and operated the system they evaluated. This dual role risks unconscious bias toward favorable interpretation. The study does not employ hypothesis testing or statistical inference; reported success rates and landing rates are descriptive statistics without confidence intervals or significance tests.

\section{Related Work}

\begin{table}[!t]
\renewcommand{\arraystretch}{1.15}
\caption{Comparison with Related Work. \emph{Hybrid}: deterministic tooling combined with an LLM. \emph{Telemetry}: previously observed data from production executions used as migration input. \emph{Setting}: industrial monorepo, industrial PRs to repository owners, or research prototype or benchmark.}
\label{tab:related_work_comparison}
\centering
\footnotesize
\setlength{\tabcolsep}{3pt}
\begin{tabular}{@{}>{\raggedright\arraybackslash}p{0.26\columnwidth} >{\raggedright\arraybackslash}p{0.22\columnwidth} >{\raggedright\arraybackslash}p{0.12\columnwidth} >{\raggedright\arraybackslash}p{0.14\columnwidth} >{\raggedright\arraybackslash}p{0.14\columnwidth}@{}}
\toprule
\textbf{Work} & \textbf{Scale} & \textbf{Hybrid} & \textbf{Telemetry} & \textbf{Setting} \\
\midrule
ClangMR, T2R~\cite{wright2013clangmr, ketkar2019typemigration} & Industrial codebases & No & No & Monorepo \\
PolyglotPiranha~\cite{ketkar2024polyglotpiranha} & 1,611 PRs & No & No & PRs \\
Nikolov et al.~\cite{nikolov-2025-migrating} & Several migrations & Yes & No & Monorepo \\
Ziftci et al.~\cite{ziftci-2025-migrating} & 39 migrations & Yes & No & Monorepo \\
SPELL~\cite{ramos2026spell} & Python library migrations & Yes & No & Research \\
MigrateLib~\cite{islam2026migratelib} & 717 applications & Yes & No & Research \\
Fandina et al.~\cite{fandina2026blindspots} & 100 COBOL programs & Yes & No & Research \\
C-to-Rust~\cite{cai2025rustmap, liu2025llmigrate, farrukh2026orbit, yan2026c2rustxw} & Programs to projects & Yes & No & Research \\
Li et al.~\cite{li2026environment} & Position paper & Proposed & No & N/A \\
Rosenfeld et al.~\cite{rosenfeld2025automatingcode} & Dataset, case study & No & No & Research \\
Benchmarks~\cite{liu2026migrationbench, li2026repomodbench, saxena2025continuous} & Benchmarks & N/A & No & Research \\
\midrule
\textbf{This work} & \textbf{2,333 PRs (729 merged)} & \textbf{Yes} & \textbf{Yes} & \textbf{PRs} \\
\bottomrule
\end{tabular}
\end{table}

Large-scale automated code migration predates LLMs. ClangMR and T2R showed that deterministic program analysis and transformation can scale to industrial codebases~\cite{wright2013clangmr,ketkar2019typemigration}, and PolyglotPiranha showed that lightweight declarative rules can express multi-language transformations, deleting 210K lines of dead code and migrating 20K lines across 1,611 pull requests in an industrial deployment~\cite{ketkar2024polyglotpiranha}. Previous work at Bloomberg has reported on successful deployments of pre-LLM automated program repair techniques into industrial development workflows~\cite{kirbas2021apr}. These systems are highly effective when a migration can be captured explicitly as rules, but become harder to apply as repository structure and migration logic become more heterogeneous.

LLMs extend this design space by handling transformations that are difficult to encode exhaustively. At Google, Nikolov et al. describe enterprise migration workflows that combine programmatic analysis and discovery, model-generated edits, compilation, testing, and human review~\cite{nikolov-2025-migrating}; Ziftci et al. report 39 migrations and 595 submitted changes using an LLM-assisted workflow~\cite{ziftci-2025-migrating}. These are the closest industrial precedents to our work. Our contribution is therefore not the combination of LLMs with traditional tooling itself, but the explicit boundary between them: classification, context construction, and verification remain deterministic, while the LLM is confined to bounded transformation and remediation steps.

Other recent systems explore related hybrid designs. MigrateLib combines LLMs with static and dynamic analysis for Python library migration~\cite{islam2026migratelib}, while SPELL uses LLMs to synthesize reusable programmatic edits~\cite{ramos2026spell}. Work on C-to-Rust migration similarly combines LLM generation with compiler, analysis, or execution feedback~\cite{cai2025rustmap,liu2025llmigrate,farrukh2026orbit,yan2026c2rustxw}, and Fandina et al. show that a domain-specific checker catches errors that LLM-only evaluation misses~\cite{fandina2026blindspots}. On the evaluation side, MigrationBench measures repository-level migration~\cite{liu2026migrationbench}, and RepoMod-Bench shows success dropping sharply as repositories grow~\cite{li2026repomodbench}; other work studies the role of context and of continuously derived benchmarks~\cite{rosenfeld2025automatingcode,saxena2025continuous}. Environment-in-the-Loop argues that executable environment state and runtime feedback should directly be a part of the migration process~\cite{li2026environment}. Our PyPackaging pipeline differs in using previously observed production telemetry (interpreter startup logs, module usage traces, and deployed dependency versions) as input to migrations.

Finally, code migration remains a human and organizational process. Prior work on human--AI partnership in LLM-based migration identifies review responsibilities and trust calibration as central to effective use~\cite{omidvartehrani-2024-evaluating}. Previous work at Bloomberg has similarly studied the developer-facing challenges of deploying automated code changes, including developers' experiences and interaction preferences~\cite{winter2022developer}, how generated changes can be delivered without creating distracting unsolicited PRs~\cite{williams2024usercentric}, and how agentic PRs should be integrated in developer workflows~\cite{williams2026pomona}. This motivates evaluating not only whether migrations can be generated, but also whether owning teams review, accept, and deploy them. Table~\ref{tab:related_work_comparison} summarizes the closest related work along the dimensions most relevant to our study: scale, hybrid architecture, runtime information, and how generated changes are delivered.

\section{Conclusions}
This paper reports on our experience using LLM-assisted pipelines to drive two large migrations at Bloomberg: a C++ build system migration and a Python packaging migration. Together they generated more than 2,300 pull requests.

The most significant change this work brought to Bloomberg was in how a central team runs a migration. Before, the team owning a migration would publish instructions and an FAQ and ask every team to do the work themselves, with each team hitting the same problems independently. Now, the central team writes those instructions for a pipeline, and the pipeline produces the PRs. Every problem solved on one repository goes into the archetype templates, prompts, and validation rules, meaning it is solved for the rest. Most of the design effort went into the deterministic parts of the pipeline, and runtime production data turned out to be an input we could not do without, although we have not yet measured its effect in a controlled way. As models improved, we gave them more of the generation work, but we kept verification deterministic.

Cheaper PRs do not make a migration free. Owning teams still have to review, test, and deploy every change, and that cost was higher where the feedback loop was deeper. For CppBuild-to-CMake, the pipeline builds every migrated project and runs its unit tests before opening a PR. In that migration, compilation is a good proxy for correctness; 34.2\% of PRs were merged and 8.1\% rejected. For PyPackaging, packaging and basic checks pass, but many Python failures only show up at runtime. Developers trusted those changes less: 10.5\% were merged and 21.8\% rejected. The two populations were also very different. The CppBuild projects fell into a few similar archetypes, so what we learned on one repository applied to the next. PyPackaging candidates spanned 18 categories with many one-off cases, so there was much less to reuse. With two campaigns, we cannot tell how much of the gap comes from feedback-loop depth and how much from the population; both played a part. In both campaigns, most PRs (67.7\% and 57.7\%) received no developer decision, reflecting the low priority of these migrations.

Our main takeaways are:
\begin{itemize}
    \item Automation makes lower-priority migrations affordable, but deciding whether a migration is worth doing is still a cost decision, and that cost includes the time owning teams spend receiving it.
    \item The greater the proportion of the feedback loop that could be run within the pipeline, the more confident the receiving teams were in the result.
    \item A homogeneous population gets far more out of this approach than one made of one-off cases.
    \item In our campaigns, the more it cost a team to receive a migration, the less likely they were to land it.
\end{itemize}

For future work, we want to allow teams to request deep-loop migrations when they are ready to deploy and verify them, rather than generating unprompted PRs. For shallow-loop migrations, where pre-PR confidence is already high, we want to explore other ways to increase engagement. These results come from one organization and two campaigns, and we would like to see them replicated elsewhere.

\section{Data Availability}
Since this project has been developed within Bloomberg’s internal systems and repositories, data and source code are proprietary and therefore cannot be made publicly accessible. However, we believe our description of the information and analysis process is sufficient for practitioners and researchers to adapt our approach in different contexts.

\bibliographystyle{IEEEtran}
\bibliography{IEEEabrv,biblio}
\end{document}